\documentclass[pre,aps,10pt,twocolumn]{revtex4-1}
\usepackage{geometry}
\usepackage{graphicx}
\usepackage{amsfonts}
\usepackage{amsmath}
\usepackage{bm}
\usepackage{url}
\usepackage{color}
\usepackage{xfrac}
\usepackage{array,url,kantlipsum}

\definecolor{purple}{rgb}{0.5, 0.0, 0.8}
\definecolor{orange}{rgb}{0.9, 0.6, 0.0}

\begin{document}
\title{Ground state of dipolar hard spheres confined in channels} 

\author{Florian Dei{\ss}enbeck}
\affiliation{Institut f\"ur Theoretische Physik II, Weiche Materie: Heinrich-Heine-Universit\"at D\"usseldorf, Universit\"atsstr.\ 1,
40225 D\"usseldorf, Germany}
\author{Hartmut L\"owen}
\affiliation{Institut f\"ur Theoretische Physik II, Weiche Materie: Heinrich-Heine-Universit\"at D\"usseldorf, Universit\"atsstr. 1,
40225 D\"usseldorf, Germany}
\author{Erdal C.~O\u{g}uz}
\affiliation{School of Mechanical Engineering and The Sackler Center for Computational Molecular and Materials Science, 
Tel Aviv University, Tel Aviv 6997801, Israel}

\date{\today}

\begin{abstract}
We investigate the ground state of a classical two-dimensional system of hard-sphere dipoles confined
between two hard walls. Using lattice sum minimization techniques we reveal that at fixed wall separations, 
a first-order transition from a vacuum to a straight one-dimensional chain of dipoles occurs upon increasing 
the density. Further increase in the density yields the stability of an undulated chain as well as nontrivial 
buckling structures. We explore the close-packed configurations of dipoles in detail, and we find that, in general, 
the densest packings of dipoles possess complex magnetizations along the principal axis of the slit. 
Our predictions serve as a guideline for experiments with granular dipolar and magnetic colloidal suspensions 
confined in slit-like channel geometry.
\end{abstract}

\maketitle 

\section{Introduction}

The anisotropy of the dipole-dipole interaction causes an ensemble of magnetic balls to self-assemble
into nontrivial ground states at zero temperature 
\cite{Tao,Groh,Dijkstra,Weis,Spiteri,Levesque}. For example, $N$ hard spheres with a
central dipole-dipole interaction exhibit ground-state clusters which cross over from linear chains to
rings and to tubes as $N$ grows \cite{Vella,Messina_2014}, whereas similar clusters have also been observed 
for anisotropically shaped magnetic particles at finite temperatures \cite{Tierno_persp,Tierno_SM2016}. 
Other ground-state structures have been explored 
for spheres with shifted dipole moments \cite{Kantorovich_2011,Kantorovich_2013,Kantorovich_2015,Yener_2016},
for spheres in an external magnetic field \cite{Messina_2017} and confined onto the plane
\cite{Kantorovich_2009,Annunziata_2013}. Magnetic hard spheres are realized in the macroworld as 
heavy balls \cite{Rehberg1} and granulates \cite{Vandewalle}, yet a plethora of possibilities are found in the mesoscopic 
regime of magnetic colloids 
\cite{Maret,Messina_onion_2015,Baraban,Weijia_2000,Erb_colloids,Alert_2017,Philipse1,Philipse2,Steinbach,Jones,Darras,Granick}, 
magnetic nanoparticles \cite{nanoparticles,Tripp_2003,Talapin} (in particular, when the magnetic interaction energy dominates thermal 
fluctuations), colloidal particles with an induced electric 
dipole moment \cite{electric1,electric2} and dipolar dusty plasmas \cite{dusty_plasmas}. 
As these particles and their clusters constitute the main building blocks of prospective materials such as ferromagnetic 
filaments \cite{Messina_fila,Goyeau_2017,Klumpp_2017} for the creation of microdevices and for magneto-rheological fluids and ferrogels with 
tunable and unusual visco-elastic properties \cite{Odenbach_2006,Cremer,Huang2016,Mughal}, an understanding of their structure is of prime interest.

In this work, we explore the ground state of classical hard sphere dipoles in two dimensions which are confined between
two narrow walls in a slit geometry (a "channel"). If the slit width coincides exactly with the hard sphere diameter,
there is only a one-dimensional degree of translational freedom for the sphere centers. In this special limit,
a vacuum coexists with a closely packed linear chain of touching dipoles with head-to-tail attractive
configuration. Here, we study the nontrivial effect of a wider slit where geometric packing effects of the disks
tends to widen the chain by buckling \cite{Schmidt_1997} but the dipolar interaction still keeps the particle chain aligned. 
As a result of this competition, we find that various {\it undulated chains\/} are the minimal potential-energy structure at 
densities slightly higher than for the straight chain. The 'spin structure' of the magnetic moments is nontrivial though mainly 
aligned with the slit. The opposite limit of high density is dictated by the close-packing problem of disks between two hard lines. 
We revisit this elementary geometric problem and analyze the spin structure in close-packed configurations, which, in general, possess 
nontrivial complex magnetizations. In principle, our predictions are verifiable in experiments with granular dipolar
particles and confined magnetic colloidal suspensions and they display relevance for the flow of magneto-rheological 
suspensions through microfluidic devices.

\begin{figure}[h!]
 \centering
 \includegraphics[width=8.0cm]{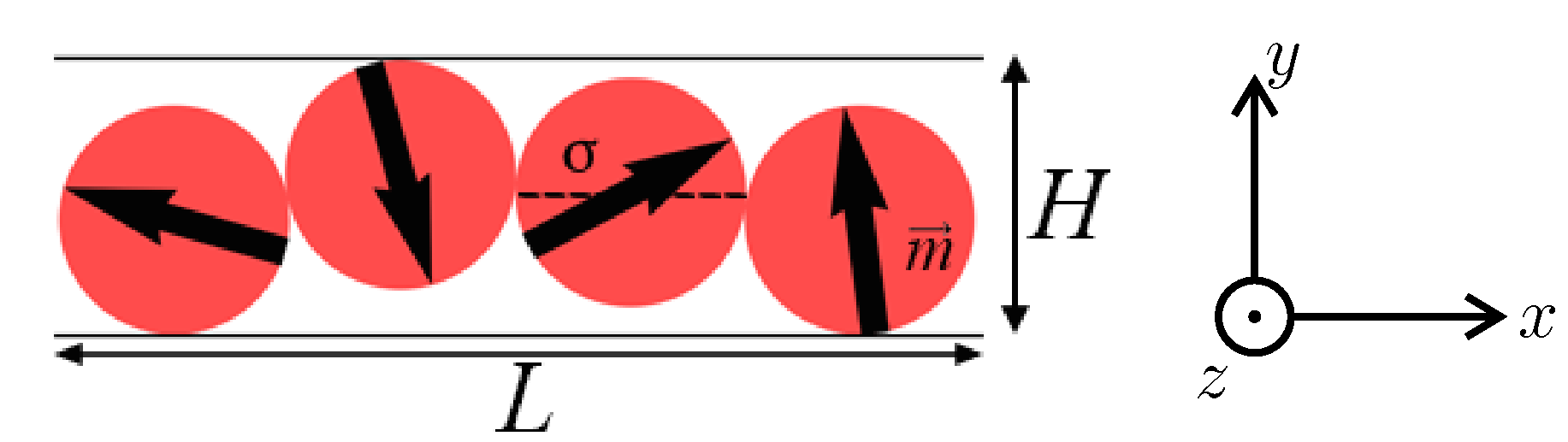}
 \caption{Schematic illustration of dipolar hard spheres of diameter $\sigma$ confined between two hard walls 
 	 of separation $H$ and length $L$. The spheres possess magnetic moments $\vec{m}$ and are further confined onto the $(x,y)$-plane.}
 \label{fig1}
\end{figure}

In our model, we consider $N$ dipolar hard spheres of diameter $\sigma$ that are confined 
in a slit-like geometry between two parallel hard walls.  
We restrict ourselves to the situation where all particles lie in the $(x,y)$-plane. This restriction yields their
magnetic moments to be coplanar with the same plane in the ground state \cite{Messina_onion_2015}, and as such, 
our system can be portrayed as an effective two-dimensional system of magnetic hard disks confined between two hard lines
of length $L$ and separation $H$, see Fig.\ \ref{fig1}. 
For convenience, this separation direction is taken along the $y$-direction. 
The potential of interaction $U(\vec{r}_{ij})$ between two constitutive particles $i$ and $j$ whose 
centers are located at $\vec{r}_i$ and $\vec{r}_j$ reads as 
\begin{equation}
 \label{eq1}
 U(\vec{r}_{ij}) = \dfrac{\mu_0}{4\pi} \left[ \dfrac{ \vec{m}_i \cdot \vec{m}_j } {r_{ij}^3}  - 
 		   3 \dfrac{ \left( \vec{m}_i \cdot \vec{r}_{ij} \right) \left( \vec{m}_j \cdot \vec{r}_{ij}\right) }{r_{ij}^5} \right]  
\end{equation}
for $r_{ij} \ge \sigma$ or infinite otherwise, where $\mu_0$ represents the vacuum permeability, and 
$r_{ij} = |\vec{r}_{ij}| = |\vec{r}_j - \vec{r}_i|$ is the interparticle distance between the spheres possessing the dipole moments $\vec{m}_i$ 
and $\vec{m}_j$ of equal magnitude $m$. At zero temperature, for a given reduced line density $\eta = N \sigma / L$, 
the system will minimize its potential energy per particle and the resulting structure will solely depend on the reduced slit width $H/\sigma$.

We employ three different approaches to study the phase behavior of our system in its entireness:
i) we theoretically determine the stability regime of undulated chains upon slight increase of the density beyond $\eta_{\mathrm{ch}}=1$ at
which the linear magnetic chain is stable, ii) we perform numerical penalty optimization method to efficiently find close-packed
structures at relatively high densities $\eta_{\mathrm{cp}}$, and iii) we carry out lattice sum minimizations to obtain the ground-state 
structures at densities interpolating between $\eta_{\mathrm{ch}}$ and $\eta_{\mathrm{cp}}$. Details of each method are provided in the following paragraphs.

\begin{figure}[!htbp]
 \centering
 \includegraphics[width=8.0cm]{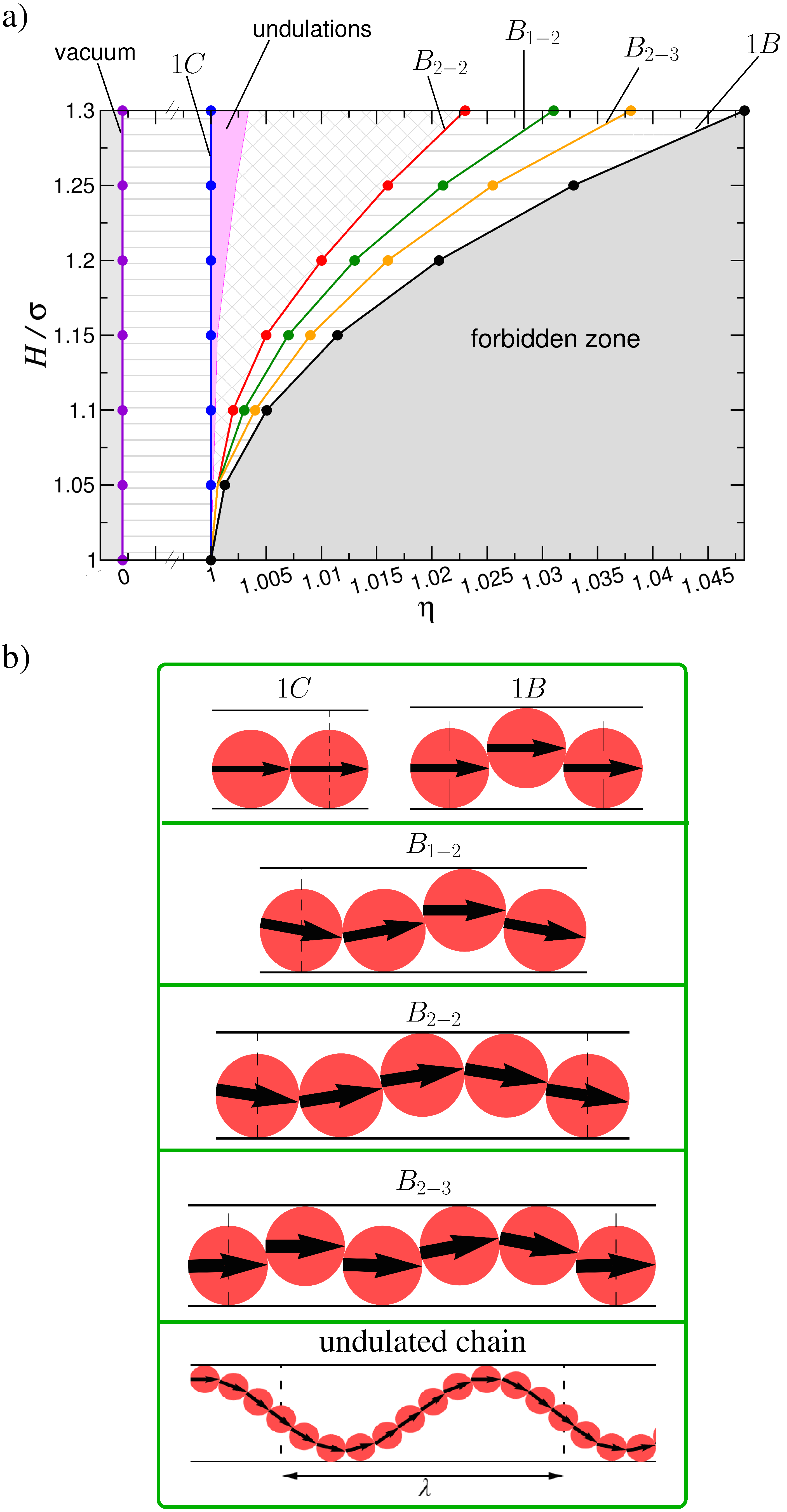}
 \caption{a) Ground-state phase diagram of confined dipolar hard spheres as a function of the reduced slit width $H/\sigma$ and the reduced 
          density $\eta$. The straight chain $1C$ (blue line) coexists with vacuum (purple line, $\eta=0$) at densities $\eta<1$, 
	  whereas for $\eta > 1$ undulated chain (purple area), the non-trivial buckling phases $B_{2-2}$ (red curve), $B_{1-2}$ (green curve),
	  and $B_{2-3}$ (yellow curve) as well as the trivial buckling phase $1B$ (black curve) become stable. The gray area indicates the forbidden  
	  zone beyond the densest packing regime. The dashed areas correspond to the coexistence regimes between two neighboring phases, whereas
	  the hatched area may accommodate further phases that have not been investigated in this work. b) Schematics of the stable phase structures  
	  shown in a). The vertical dashed lines mark the corresponding unit cells, and $\lambda$ in the most lower schematic indicates the wavelength 
	  of the undulated chain with a sinusoidal form. }
 \label{fig2}
\end{figure}
We wish to describe first our lattice-sum technique. At each prescribed reduced density $\eta$ and reduced slit width $H/\sigma$, 
we perform lattice sum minimizations for a broad set of candidate structures, which we take to be two-dimensional crystals with periodicity 
along the $x$-direction, and confined along the $y$-direction by the slit, cf.\ Fig.\ \ref{fig1}.  
We consider structures with a rectangular primitive cell containing up to $n=6$ particles. Without any further restriction, we minimize 
the energy per particle with respect to the particle coordinates in the cell and the alignments of the two-dimensional dipole moments.
The resulting ground-state phase diagram in $(\eta,H/\sigma)$-plane is demonstrated in Fig.\ \ref{fig2}a for $0 \le \eta \lesssim 1.05$ 
and  $1 \le H/\sigma \le 1.3$.

The phase diagram exhibits a relatively large coexistence regime between the vacuum at $\eta = 0$ (purple line in 
Fig.\ \ref{fig2}a) and the linear chain of touching particles with head-to-tail attractive dipole configurations at the 
density $\eta_{\mathrm{ch}} = 1$ (blue line in Fig.\ \ref{fig2}a). The linear chain $1C$ is schematically depicted in Fig.\ \ref{fig2}b. 
Upon increasing the density beyond $\eta_{\mathrm{ch}}$, we observe the stability 
of undulated chains within the purple area shown in Fig.\ \ref{fig2}a. The undulations of the magnetic 
chain are caused by an interplay of the geometry that favors a buckled chain 
for slit widths $H/\sigma > 1$ and densities $\eta > \eta_{\mathrm{ch}}$ due to an efficient packing and 
the attractive head-to-tail alignment of the dipoles favoring the straight chain. 
To investigate the stability of the undulated chain, we have calculated its total potential energy by taking into account its bending energy. 
Since lattice-sum minimization techniques are inappropriate to determine the energy of highly complex undulated phase structures with a large number 
of particles per unit cell that even diverge as $\eta \to \eta_{\mathrm{ch}}$ for $\eta > \eta_{\mathrm{ch}}$, we have used a different approach 
to detect such stable undulations as will be described in the following.

We model the undulated chain as a continuous object uniformly carrying dipolar hard spheres. We assume that the linear chain 
undulates into periodic sinusoidal structures with a density-dependent wavelength $\lambda$ as schematically illustrated in Fig.\ \ref{fig2}b. Thus, 
we parametrize the contour line of the wave that passes through the particle centers by $f(x) = \frac{H-\sigma}{2} \cos{(kx)}$, with $k=2\pi/\lambda$. 
The wavelength $\lambda$ relates to the density via $\eta = l_{c} / \lambda$, where 
$l_{c}= \int_0^{\lambda} \sqrt{1+f^{\prime}(x)^2} \mathrm{dx}$ denotes the contour length of the corresponding wave.
Inserting $f(x)$ into $l_c$ yields the equality $2\pi\eta - E(2\pi,-k(H-\sigma)/2) = 0$, from which we obtain $k$ (and thus $\lambda$) for a given 
reduced density $\eta$. Here, $E$ stands for the incomplete elliptic integral of the second kind. 

Having established the exact form of the undulation wave with the corresponding wavelength, we will now determine its potential energy
per particle, $u_{\mathrm{und}}$, given as
\begin{equation}
 \label{eq2}
 u_{\mathrm{und}} = u_{\mathrm{ch}} + \dfrac{\gamma}{N} \int_0^{l_{c}} \dfrac{1}{R(s)^2} \mathrm{ds}.
\end{equation}
The first term on the right hand side, $u_{\mathrm{ch}}$, describes the magnetic energy per particle in the straight chain as dictated by 
the pair interaction potential of Eq.\ \ref{eq1}, and the second term the elastic bending energy per particle, respectively. 
$R(s)$ denotes the curvature radius of the wave as a function of its arc length 
$s(x) = \int_0^x \sqrt{1+f^{\prime}(x^{\prime})^2} \mathrm{dx^{\prime}}$, whereas $\gamma$ stands for the bending rigidity of the magnetic chain 
as will be determined in the following.

To obtain the bending rigidity $\gamma$, we follow the approach proposed in Ref.\ \cite{Vella}; namely, we calculate $\gamma$ by identifying the bending
energy with the energy difference of a closed ring of radius $R$ carrying $N$ touching particles and a straight infinite chain.
We consider the magnetic moments in the ring configuration to be oriented tangentially on the bending circle. 
The tangential alignment has been theoretically shown to minimize the ring's potential energy per particle $u_{\mathrm{ring}}$ in \cite{Kiani_2015}, 
and it has been demonstrated experimentally with cobalt nanoparticles in small systems in \cite{Tripp_2003}. 
Using the relation $u_{\mathrm{ring}}(R) - u_{\mathrm{ch}} \sim \gamma / R^2 + O(R^{-4})$, and ignoring the higher order terms in $R$, the bending 
rigidity $\gamma$ is given as 
\begin{equation}
 \label{eq3}
 \gamma = \lim_{R \to \infty} [ u_{\mathrm{ring}}(R) - u_{\mathrm{ch}} ] R^2 . 
\end{equation}
Taking into account the closed-form expressions for the energies $u_{\mathrm{ring}}$ and $u_{\mathrm{ch}}$ (cf.\ \cite{Vella,Messina_2014, Messina_onion_2015}), 
we finally obtain $\gamma 4\pi \sigma^2 / \mu_0 m^2 = (\zeta (3) +1/6)/4 \approx 0.342$, where $\zeta (m) = \sum_{i=1}^{\infty} i^{-m}$ denotes 
the Riemann zeta function.

We calculate the total energy per particle of the undulated chain as given in Eq.\ \ref{eq2}, and compare it to the energies obtained by our 
lattice-sum minimizations. As a result, we reveal the stability of the undulated chain for slightly larger densities above
$\eta_{\mathrm{ch}} = 1$ and for all slit widths. 
In the limit $\eta \to \eta_{\mathrm{ch}}$ with $\eta > \eta_{\mathrm{ch}}$, the wavelength of undulations diverge. 
As the reduced density increases, undulations become stable with continuously decreasing wavelengths.
For instance, at $H/\sigma = 1.3$ the smallest wavelength we obtain is $\lambda \approx 16\sigma$. 

The phase space between the undulated chain and the trivial buckling structure $1B$ at the close-packing density $\eta_{\mathrm{cp}}$ 
(black curve in Fig.\ \ref{fig2}a) displays a plethora of non-trivial buckling structures as obtained by lattice-sum minimizations. 
We refer to these phases as $B_{2-2}$ ($n=4$), $B_{1-2}$ ($n=3$), and $B_{2-3}$ ($n=5$) and we show their stability regime by the red, green, 
and yellow curves in Fig.\ \ref{fig2}a, respectively. The subindices indicate the number of particles per unit cell that are distributed 
--not necessarily evenly-- on the two walls while being in contact with them. For instance, $B_{2-3}$ possesses five primitive-cell particles, 
where two of them are in contact with the one, and three with the other wall.

For the sake of completeness, we have further investigated the phase coexistence in our system by implementing the 
common tangent (Maxwell) construction: The dashed areas between two neighboring phases in Fig.\ \ref{fig2}a demonstrate the their coexistence regime. 
As a result, we obtain the pure one-phase stability of the non-trivial buckling phases at a single density for a given slit width, 
and as such, they emerge as stability lines in the $(\eta,H/\sigma)$-plane.

Attention must be paid when interpreting the hatched area in Fig.\ \ref{fig2}a. On the one hand, it might indicate a coexistence between 
the undulated phases and $B_{2-2}$, on the other hand further complex phases that have not been investigated in this work may occur within this area: 
So far, we have considered a single mode of undulation of the linear magnetic chain, namely periodic sinusoidal waves functions. 
In principle, other types of undulations might take place which, if stable, are expected to appear within this hatched area. Moreover, phase structures 
with more than 6 unit cell particles cannot be ultimately excluded. This being said, however, we do not expect any radical morphology changes
of the phase diagram.

The close-packing density $\eta_{\mathrm{cp}}$ shown by the black curve in Fig.\ \ref{fig2}a is an upper bound of the phase space. 
The gray area beyond this density displays the geometrically inaccessible ('forbidden') density zone. In order to reveal the minimum-energy state 
of hard dipolar disks along this density, we first use the \textit{penalty} method to find the maximum-packing configuration of 
hard disks. This method as implemented in \cite{ErdalPRL_2012,Assoud_2011} describes an efficient algorithm
to circumvent the discontinuous and constrained optimization of the free space under the constraint of non-overlapping particles.
By adding a penalty term that depends continuously on the overlap area of two disks, we obtain a continuous and unconstrained
penalty function which can be minimized in the classic way to predict the optimal particle coordinates.
Subsequently, in a given densest packing structure, we first assign magnetic moments to each disk and we then minimize
the potential energy with respect to the alignments of those moments by our lattice sum minimization technique. 
As a result, we unveil the trivial buckling structure $1B$ with $n=2$, where the dipole moments are all aligned parallel to the $x$-axis, 
cf.\ the corresponding schematic in Fig.\ \ref{fig2}b.

\begin{figure}[h!]
 \centering
 \includegraphics[width=8.5cm]{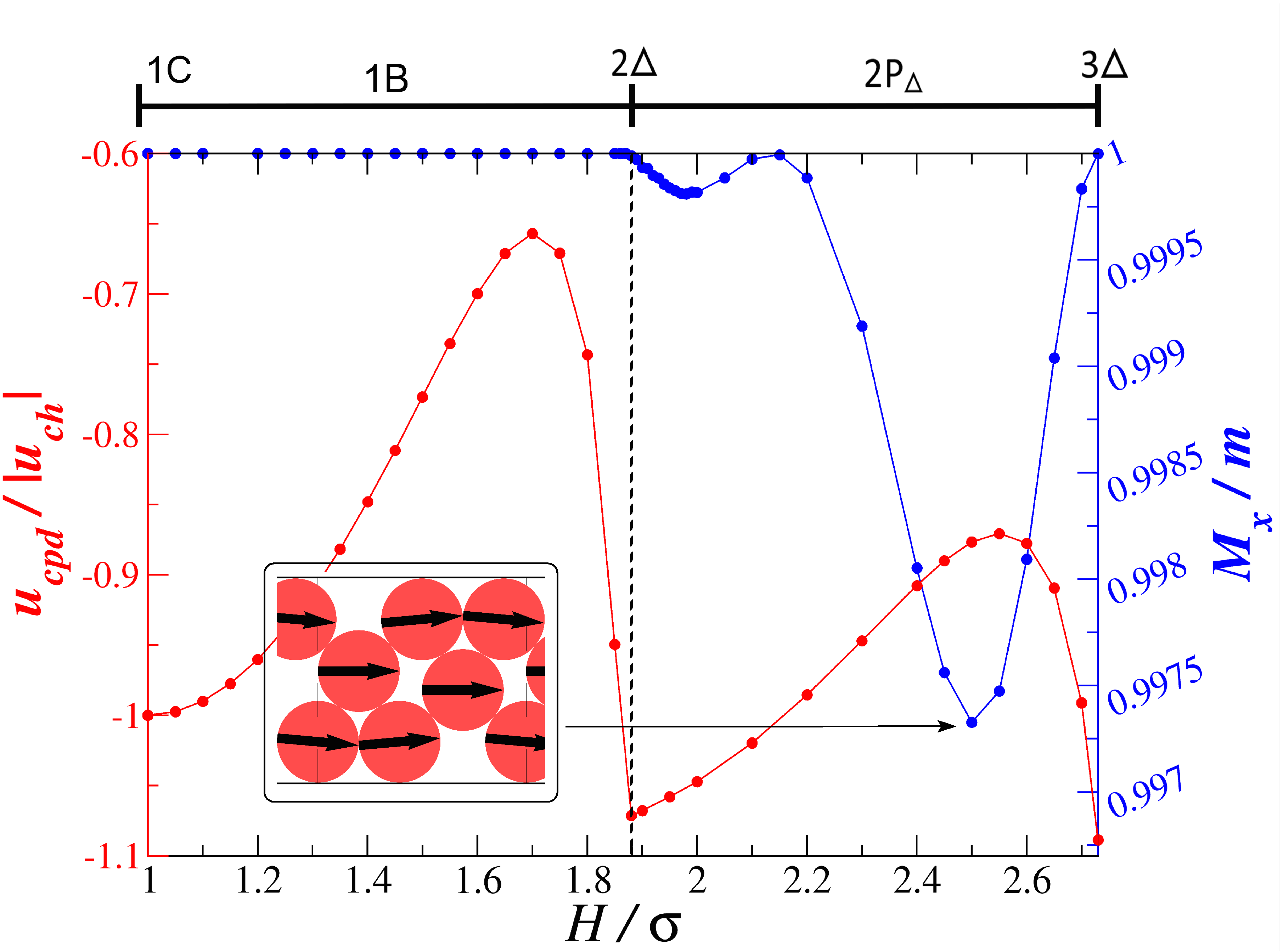}
 \caption{Potential energy per particle of closed-packed dipolar hard spheres (left axis) and their total magnetization per particle in the parallel  
 	  direction to the walls (right axis). The corresponding close-packed structures are shown in the upper panel. The inset provides a schematic 
	  illustration of dipolar $2P_{\triangle}$ at $H/\sigma = 2.5$.}
 \label{fig3}
\end{figure}

Next, we wish to examine in detail the magnetic spin structure, i.e., the orientations of the magnetic moments, of close-packed hard disks 
for larger slit widths. To this end, we first extend our geometrical study of the densest packings up to $H/\sigma \le \sqrt{3} + 1$. 
In particular, we investigate the regime between the linear chain $1C$ at $H/\sigma = 1$ and the triangular trilayer $3\triangle$ at 
$H/\sigma = \sqrt{3} + 1$.
We obtain the fundamental sequence $w\triangle \to (w+1)\triangle$ as well as the intermediate phases $1B$ and $2P_{\triangle}$ as reported 
in \cite{Molnar_1978,Furedi_1991}, where $w \ge 2$ denotes the number of parallelly stacked and staggered chains aligned with the walls, 
corresponding to slices of the regular triangular lattice. The phases $w\triangle$ are only best packed at discrete values of the slit width, 
$H_w/\sigma = (w-1) \sqrt{3}/2 + 1$, where the layers exactly fit between the walls.
The resulting cascade of close-packed structures reads as $1C \to 1B \to 2\triangle \to 2P_{\triangle} \to 3\triangle$, 
and it is shown in the upper panel of Fig.\ \ref{fig3}.

Having elucidated the close-packed structures of confined hard disks, we identify their energy-minimizing spin configurations using our 
lattice-sum method. The resulting potential energy per particle $u_{cpd}$ of close-packed dipolar hard spheres is shown in Fig.\ \ref{fig3}. 
We further calculate the total magnetization per particle along the $x$-direction by
\begin{equation}
M_x = \dfrac{1}{n} \sum_{i=1}^n m \cos{\theta_i}, \\
\label{eq4}
\end{equation}
where $\theta_i$ denotes the angle between the magnetic moment $\vec{m}_i$ and the $x$-axis. In Fig.\ \ref{fig3},  
$M_x$ is plotted for different densest packings as a function of $H/\sigma$. We observe that the phases $2\triangle$ and $3\triangle$
as well as the trivial buckling phase $1B$ exhibit $M_x=m$ as their moments are all aligned parallel 
to the $x$-axis. The phase $2P_{\triangle}$ possesses, however, a non-trivial structure of the magnetic moments that are rotated with 
respect to the $x$-axis as shown in the inset of Fig.\ \ref{fig3} for $H/\sigma = 2.5$. Consequently, the total magnetization $M_x$ differs
slightly from the magnetization $m$ of a straight chain.

\section{Conclusions}

In conclusion we have explored the ground-state structures of strongly confined magnetic disks
in a slit geometry and have predicted a novel structure of undulated chains which emerge as a compromise between packing efficiency 
and magnetic dipole moment alignment. This simple model system could be realized either in the granular or in the 
colloidal context. For colloids, the two dimensionality of our model is a standard set-up, and the slit geometry can be imposed 
by microchannels (see e.g. \cite{Nielaba}) or by strong external fields. Another promising direction is the structure 
formation of colloids under adaptive confinement where the corresponding boundaries are made of a subset of constitutive particles fixed with optical tweezers 
\cite{Erdal_natp2016,Erdal_jcp2014}.  
Furthermore, for the future, an ensemble of active or 
swimming magnetic particles in a slit would be a fascinating topic where the emerging clusters are not static but dynamic \cite{active1,active2}. 
Moreover more general models including an external magnetic field 
\cite{external} or  harmonic confining potential and an out-of-plane orientation of the dipole moments can be explored using similar techniques as presented here.
Finally the fact that magnetization can be tuned
 paves the way for microdevices to control the magnetization 
intrinsically by design architecture.

\begin{acknowledgements}
We thank Frank Smallenburg for fruitful discussions. Financial 
support by the Deutsche Forschungsgemeinschaft (DFG) within project LO 418/19-1
is acknowledged.
\end{acknowledgements}



\end{document}